\documentclass[runningheads]{llncs}
\usepackage[T1]{fontenc}

\usepackage{graphicx}
\usepackage{booktabs}
\usepackage{hyperref} 
\usepackage[misc]{ifsym}
\begin{document}
\title{Potential Indicator for Continuous Emotion Arousal by Dynamic Neural Synchrony}
\titlerunning{Potential Indicator for Continuous Emotion Arousal}
%

\author{Guandong Pan\inst{1, 3} \and
Zhaobang Wu\inst{1, 3}  \and
Yaqian Yang\inst{2, 3}\and
Xin Wang\inst{2, 3, 5, 6 ,7}  \and
Longzhao Liu\inst{2, 3, 5, 6 ,7}  \and
Zhiming Zheng\inst{2,3,4,5,6,7,8} \and
Shaoting Tang \inst{2,3,4,5,6,7,8}$\textsuperscript{(\Letter)}$  }
\authorrunning{G. Pan et al.}
%
\institute{School of Computer Science and Engineering, Beihang University, Beijing 100191, China \and
Institute of Artificial Intelligence, Beihang University, Beijing 100191, China \\
\email{zzheng@pku.edu.cn} \and
Key laboratory of Mathematics, Informatics and Behavioral Semantics, Beihang University, Beijing 100191, China \\
\email{\{pan\_gd, wuzhaobang, yangyaqian, wangxin\_1993, longzhao, tangshaoting\}@buaa.edu.cn} \and
Institute of Medical Artificial Intelligence, Binzhou Medical University, Yantai 264003, China  \and
Zhongguancun Laboratory, Beijing 100094, China \and
Beijing Advanced Innovation Center for Future Blockchain and Privacy Computing, Beihang University, Beijing 100191, China  \and
PengCheng Laboratory, Shenzhen 518055, China  \and
State Key Lab of Software Development Environment, Beihang University, Beijing 100191, China}
\maketitle              
\begin{abstract}
The need for automatic and high-quality emotion annotation is paramount in applications such as continuous emotion recognition and video highlight detection, yet achieving this through manual human annotations is challenging. Inspired by inter-subject correlation (ISC) utilized in neuroscience, this study introduces a novel Electroencephalography (EEG) based ISC methodology that leverages a single-electrode and feature-based dynamic approach. Our contributions are three folds: Firstly, we reidentify two potent emotion features suitable for classifying emotions—first-order difference (FD) an differential entropy (DE). Secondly, through the use of overall correlation analysis, we demonstrate the heterogeneous synchronized performance of electrodes. This performance aligns with neural emotion patterns established in prior studies, thus validating the effectiveness of our approach. Thirdly, by employing a sliding window correlation technique, we showcase the significant consistency of dynamic ISCs across various features or key electrodes in each analyzed film clip. Our findings indicate the method’s reliability in capturing consistent, dynamic shared neural synchrony among individuals, triggered by evocative film stimuli. This underscores the potential of our approach to serve as an indicator of continuous human emotion arousal. The implications of this research are significant for advancements in affective computing and the broader neuroscience field, suggesting a streamlined and effective tool for emotion analysis in real-world applications.

\end{abstract}


\section{Introduction}

Capturing the ground truth of dynamic human emotions is pivotal for advancements in continuous emotion recognition and video highlight detection \cite{baveye_liris-accede_2015}, \cite{chenes_highlight_2013},  \cite{ding_inter-brain_2021}. Traditionally, the collection of two-dimensional emotional responses, namely arousal and valence, has relied on subjective human annotations \cite{ding_inter-brain_2021}, \cite{soleymani_analysis_2016}. However, the dual demands of video observation and annotation tasks impose a cognitive burden on annotators, which can compromise the quality of the data collected. Additionally, the requirement for conscious rating in such tasks may affect long-term efficiency. These limitations pose a significant scientific challenge: Is there an alternative method to annotate human emotions both unconsciously and efficiently?

The technique of inter-subject correlation (ISC), derived from neuroscience, offers a promising approach to capturing group-level shared neural responses without conscious effort from participants \cite{hasson_reliability_2010}, \cite{hasson_intersubject_2004}, \cite{hasson_hierarchy_2008},  \cite{nastase_measuring_2019}. This neural synchrony, reflecting the processing of emotional information in response to evocative video content \cite{sachs_dynamic_2020}, serves as an indicator of variations in human emotional arousal. Traditional ISC research, primarily utilizing functional Magnetic Resonance Imaging (fMRI), has established the reliability of this method across different cinematic experiences \cite{hasson_reliability_2010}. In contrast, electroencephalography (EEG) offers advantages over fMRI in terms of affordability, portability, and high temporal resolution, making it more applicable to real-world tasks \cite{wei-long_zheng_investigating_2015}. Recent studies have applied EEG to extract ISC, but have predominantly focused on multichannel analyses  \cite{dmochowski_audience_2014}, \cite{dmochowski_extracting_2018}, \cite{dmochowski_correlated_2012}, which do not meet the practical needs of applications requiring fewer channels. To address this limitation, we propose a novel feature-based dynamic EEG-based ISC method computed on each individual channel. This approach can autonomously capture the continuous synchronization of brain regions among a population, thereby providing a potential neural-based indicators for representing the ground truth of human dynamic emotion arousal.

In brief, the contributions of our work are as follows: 
\begin{itemize}
    \item We reidentify two robust emotion features that are pivotal in distinguishing emotional states. The first feature, derived from the time domain, is the first-order difference (FD), and the second, from the frequency domain, is differential entropy (DE).
    \item Utilizing an overall correlation analysis, this study demonstrates the heterogeneous synchronized performance across electrodes. This finding is in alignment with established neural emotion patterns from previous research, thus serving as a validation of our methodology’s effectiveness.
    \item By employing a sliding window correlation technique, we illustrate the consistency of dynamic ISCs across various features or key electrodes within each analyzed film clip. This consistency underscores the reliability of our method in capturing dynamic shared neural synchrony among individuals, triggered by emotionally evocative film stimuli. Our approach suggests significant potential to serve as an indicator of continuous human emotion arousal. 
\end{itemize}

\section{Related Works}

\subsection{Inter-subject correlation} \label{sec:related_work_ISC}

Inter-subject correlation (ISC) quantifies the similarity in brain responses among individuals exposed to \textbf{the same naturalistic stimuli}, such as films or stories.
In the seminal work \cite{hasson_intersubject_2004}, Hasson et al. pioneered the exploration of ISC by demonstrating synchronous neural activity across participants during film viewing, not only in primary sensory areas but also in higher associative cortices involved in complex cognitive functions such as face recognition. Over the past decades, the intuitive underlying principles of ISC have facilitated its adoption across a wide range of fields, yielding significant insights into mechanisms of attention \cite{ki_attention_2016}, \cite{poulsen_eeg_2017}, \cite{rosenkranz_eeg-based_2021}, friendship prediction \cite{parkinson_similar_2018}, emotion  \cite{ding_inter-brain_2021}, \cite{hu_similar_2022}, \cite{nummenmaa_emotions_2012}, marketing \cite{barnett_ticket_2017}, \cite{dmochowski_audience_2014}, memory processes \cite{chen_shared_2017}, \cite{hasson_enhanced_2008}, video highlight detection \cite{chenes_highlight_2013}, healthcare applications \cite{hasson_shared_2009}, and even interspecies comparisons \cite{mantini_interspecies_2012}. 

Historically, the majority of ISC research utilized functional Magnetic Resonance Imaging (fMRI). However, there has been a gradual shift towards incorporating alternative modalities such as Electroencephalography (EEG) \cite{ding_inter-brain_2021}, \cite{dmochowski_audience_2014}, \cite{liu_what_2016}, Electrocorticography (ECoG) \cite{haufe_elucidating_2018}, Magnetoencephalography (MEG) \cite{lankinen_intersubject_2014}, functional Near-Infrared Spectroscopy (fNIRS) \cite{liu_measuring_2017}, Electrocardiography (ECG) \cite{perez_conscious_2021}, analysis of eye movements \cite{hasson_hierarchy_2008}, and other physiological signals \cite{chenes_highlight_2013}. Notably, Haufe et al. \cite{haufe_elucidating_2018} confirmed that fMRI, ECoG, and EEG exhibit comparable repeat-reliability across viewings, highlighting EEG’s potential for capturing shared neural responses with its high temporal resolution and suitability for naturalistic settings due to portable equipment.

Nevertheless, a significant gap in the literature exists regarding the practical implementation of ISC using EEG in real-world environments, where the use of a minimal number of electrodes is preferable. While past studies often employed multiple electrodes to enhance analysis performance \cite{dmochowski_audience_2014}, \cite{dmochowski_correlated_2012}, \cite{ki_attention_2016}, \cite{rosenkranz_eeg-based_2021}, this approach is less feasible in the real world, potentially limiting the broader application of EEG-based ISC techniques. Addressing this gap by developing methodologies for ISC analysis with fewer electrodes could significantly expand the utility and accessibility of EEG for real-world applications. By isolating the ISC analysis to each individual electrode, we leverage both overall and sliding window correlation techniques to meticulously assess the potential of single-channel data to capture neural synchrony. Results demonstrate the reliability of our methodology in capturing dynamic shared neural synchrony, suggesting the potential to serve as a reliable indicator of continuous emotional arousal. This approach not only simplifies the hardware requirements for real-world applications but also maintains analytical precision, thus significantly advancing the practical deployment of EEG-based ISC techniques.

\subsection{EEG feature extraction} \label{sec:feature}
Traditionally, inter-subject correlation (ISC) techniques have been applied directly to raw brain signals. However, in the realm of EEG-based emotion recognition, a variety of robust feature extraction methods have been developed. These methods have proven effective in distinguishing emotional valences through model training and testing. This presents a compelling case for implementing feature-related ISC analysis.

In 2013, the Differential Entropy (DE) feature was introduced \cite{duan_differential_2013}, with subsequent studies affirming its efficacy in representing emotional states \cite{wei-long_zheng_investigating_2015}, \cite{yi_learning_2023}, \cite{zheng_identifying_2019}. A comprehensive review in 2019 \cite{yu_review_2019} evaluated EEG features across four domains—time, frequency, time-frequency, and spatial—using the sparse linear discriminant analysis (SLDA) method across three distinct datasets (SEED, DREAMER, CAS-THU). This review highlighted that time-domain features, particularly the first-order difference (FD), exhibited superior performance in discriminating emotional valence.

Motivated by these findings, our study begins with a user-dependent comparative analysis of features to ascertain their efficacy in extracting emotional information. Following this preliminary evaluation, DE and FD features are identified as particularly potent and are subsequently selected for further analysis in extracting neural synchrony. This approach aims to enhance the applicability of EEG-based single-channel ISC techniques by leveraging sophisticated feature-based analyses.


\section{Materials and Methodology}
\subsection{Dataset and preprocessing}
The SEED EEG dataset, a publicly available affective dataset, was introduced by Zheng et al. \cite{wei-long_zheng_investigating_2015}. It comprises EEG recordings from fifteen Chinese participants (seven males and eight females; mean age: 23.27, SD: 2.37) who viewed various Chinese film clips designed to evoke distinct emotional valences: positive, negative, and neutral. Each participant was exposed to the same 15 clips, approximately four minutes each, across three separate sessions on different days. EEG data were captured using a 62-channel setup conforming to the International 10-20 system via the ESI Neuroscan system. For our study, 45 sessions (15 participants, three sessions each) of EEG records are analyzed to extract statistical shared neural responses at the population level for each film clip.

Our analysis utilize the "Preprocessed Data" set, sampled at 200 Hz and bandpass filtered from 0 to 75 Hz. Using MATLAB’s EEGLAB toolbox \cite{delorme_eeglab_2004}, we apply a notch filter between 48 and 52 Hz to eliminate electrical noise. Visual inspections of channel data for each session are used for the interpolation of faulty channels. Additionally, artifacts from eye movements and muscle contractions are isolated and removed through independent component analysis (ICA) implemented in EEGLAB.

\subsection{EEG-based ISC}

\subsubsection{Feature extraction}

Original features in EEG analysis are derived from the absolute raw values of EEG signals following preprocessing, whereas first-order difference features represent the differences between consecutive sample values. First-order difference features are particularly valuable as they capture the non-linear dynamics of EEG signals, namely the rate of change in voltage, which is closely linked to emotional states \cite{yu_review_2019}. In our ISC methodology, we introduce a variable, $scale$, which aggregates $s$ sample points into a single feature point. This scaling allows for the analysis of shared cognitive information across varying temporal resolutions and reduces the length of the feature vectors, thereby decreasing storage requirements and enhancing computational efficiency.

The equations for the original and first-order difference feature vectors for a single EEG record are defined as \cite{picard_toward_2001}:
\begin{equation}
    V_{O} (i) = \frac{1}{s}\sum_{j=i \cdot s}^{(i+1)\cdot s - 1} \left|x(j)\right|
, 0 \leq i <  \lfloor\frac{N}{s}\rfloor,
\end{equation}
\begin{equation}
    V_{FD} (i) = \frac{1}{s}\sum_{j=i \cdot s}^{(i+1)\cdot s - 1} \left|x(j+1) - x(j)\right|
, 0 \leq i <  \lfloor\frac{N}{s}\rfloor,
\end{equation}
where $i$ represents the index of the feature vector, $x(j)$ deontes the $j$-th sample point in the EEG record, $N$ is the total number of samples, and $s$ is the scale factor converting $s$ sample points into one feature point.

Additionally, we incorporate the Differential Entropy (DE) feature \cite{duan_differential_2013}, which provides a quantification of the complexity or uncertainty associated with EEG signal frequency distributions. It has been found that the subbands of EEG signals are nearly subject to Gaussian distribution \cite{duan_differential_2013}. The DE feature is calculated over a fixed scale of 200 sample points, corresponding to a 1-second time scale at a 200 Hz sampling rate. We compute DE features using Fast Fourier Transform with a 1-s-long window and no overlapping window. Given that a series of a certain band $X$ follows a Gaussian distribution $N(\mu,\sigma^2)$, where $\mu$ is the mean and $\sigma^2$ is the variance of the distribution, the DE is defined by the formula:
\begin{equation}
V_{DE}(i) = -\sum_{k=0}^{K-1} p(X_k) \log p(X_k) =  \frac{1}{2} \log(2\pi e \sigma^2), 0 \leq i <  \lfloor\frac{N}{s}\rfloor
\end{equation}
where $i$ represents the index of the feature vector, $p(x)$ represents the probability distribution of the EEG frequency amplitude values within the defined window, namely Gaussian distribution. This fixed-scale analysis provides a consistent measure of entropy across all EEG channels, offering insights into the informational content of the signals related to emotions. In our work, we compute DE features for four frequency bands respectively, i.e., $\delta$ and $\theta$ band (1-7Hz), $\alpha$ band (8-13Hz), $\beta$ band (14-29Hz), $\gamma$ band (30-47Hz).

\subsubsection{Similarity measurement}
Pearson correlation coefficient (PCC) is used in ISC similarity measurement that was proposed by Karl Pearson. PCC can measure a linear association or dependence, between two continuous variables \cite{nastase_measuring_2019}. Given two random variables $X,Y$ and their paired N samples $x(i), y(i), i=1,2,3,...N$, the formula of PCC is as follows:

\begin{equation}
PCC (X, Y) = \frac{Cov(X, Y)}{\sigma_X \cdot \sigma_Y} = \sum_{i=1}^n \frac{ (x_i-\bar x)(y_i-\bar y)}{\sqrt{(x_i- \bar x)^2} \sqrt{(y_i- \bar y)^2}}
\end{equation}
where $Cov(X, Y)$ means the covariance between $X, Y$, $\bar x$ and $\sigma_x$ are the mean and standard deviation of $x(i)$, $\bar y$ and $\sigma_y$ are the mean and standard deviation of $y(i)$ and $N$ means the number of samples.

\subsubsection{Overall and sliding window correlation}

Inter-subject correlation (ISC) can be computed using two distinct methodologies: overall correlation and sliding window correlation. The overall correlation method computes a single correlation coefficient, reflecting the global similarity across the entire lengths of two feature vectors. In contrast, the sliding window correlation method segments the feature vectors into smaller, overlapping windows. For each window, a separate correlation is computed, resulting in a sequence that illustrates temporal fluctuations in correlation, providing insights into dynamic synchronization patterns. The length of this sequence is influenced by both the width of the window and the overlap between successive windows, which is set to 1 second in this study.

The choice of window width is crucial and remains a subject of ongoing research \cite{mokhtari_sliding_2019}. In our analysis, window widths of 10 seconds and 70 seconds are utilized to adequately capture brief scenes and longer narrative arcs within the films, respectively. These correlation calculations are performed for each pairwise combination of subjects, across all channels and all films, resulting in a substantial dataset comprising 990 pairs, 62 channels, and 15 films within the SEED dataset. For each film and channel, sliding window correlations are averaged across all pairs of subjects to determine population-level synchronous responses. These correlations are denoted as SW-O-ISC, SW-FD-ISC, and SW-DE-ISC for the original, first-order difference (FD), and differential entropy (DE) feature vectors, respectively. To enhance computational efficiency, the Python library 'taichi' \cite{hu_taichi_2019} is employed for parallel computation of sliding window ISC.

The mathematical representations for the overall ($\zeta$) and sliding window ($\xi$) correlations are given by:

\begin{equation}
    \zeta_{i,j}^{f,c} = PCC(V_i^{f,c},V_j^{f,c}), 
\end{equation}

\begin{equation}
    \xi^{f,c} (k) = \frac{1}{M^*}\sum_{i,j} PCC(v_{i}^{f,c}(k), v_{j}^{f,c}(k)),
\end{equation}
where $i,j$ represent participant indices ($1 \leq i \leq M, i < j \leq M$, with $M$ being the total number of participants), $f,c$  denote film and channel indices respectively,. $v_{i}^{f,c}(k)$ represents the segmented feature vector in the $k$th window. $M^*$  signifies the total number of pairwise comparisons.

\section{Experiments and Results}
\subsection{EEG feature classification and selection for effective emotion representation}

\begin{table} 
\centering
  \begin{tabular}{ccccccc}
    \toprule
    Feature  & Original & FD & Hjorth & NSI & PSD &  DE  \\
    \midrule
      Acc  & 66.18 & 87.96 & 83.27 & 47.09 & 77.67 & 82.46 \\
    \bottomrule
  \end{tabular}
 	\setlength{\abovecaptionskip}{0.4cm}
  \caption{\textbf{User-dependent recognition accuracy(\%) of different feature extraction methods on SEED.} }
  \label{tab:FeatureExtraction}
\end{table}

In the field of EEG-based emotion recognition, features in the frequency domain have been widely used, such as power spectral density (PSD) and differential entropy (DE) \cite{alarcao_emotions_2019}, \cite{jenke_feature_2014}, 
\cite{wei-long_zheng_investigating_2015}, \cite{zheng_identifying_2019},. However, a recent study \cite{yu_review_2019} have showed that features in the time domain, such as the first difference, exhibit superior discriminative power in differentiating emotional valence compared to their frequency domain counterparts. 

To further explore the discriminative capabilities of these features, we conducted an experiment using the SEED dataset, employing the linear kernel Support Vector Machine (SVM) classifier. We select four time-domain features for evaluation: the original feature, the first-order difference (FD) feature, the Hjorth parameter, and the non-stable index (NSI). Additionally, we include PSD and DE from the frequency domain for comparative analysis. A detailed description of these feature extraction methods is available in \cite{yu_review_2019}. 

\begin{figure}[htbp]
  \centering
  \includegraphics[width=\linewidth]{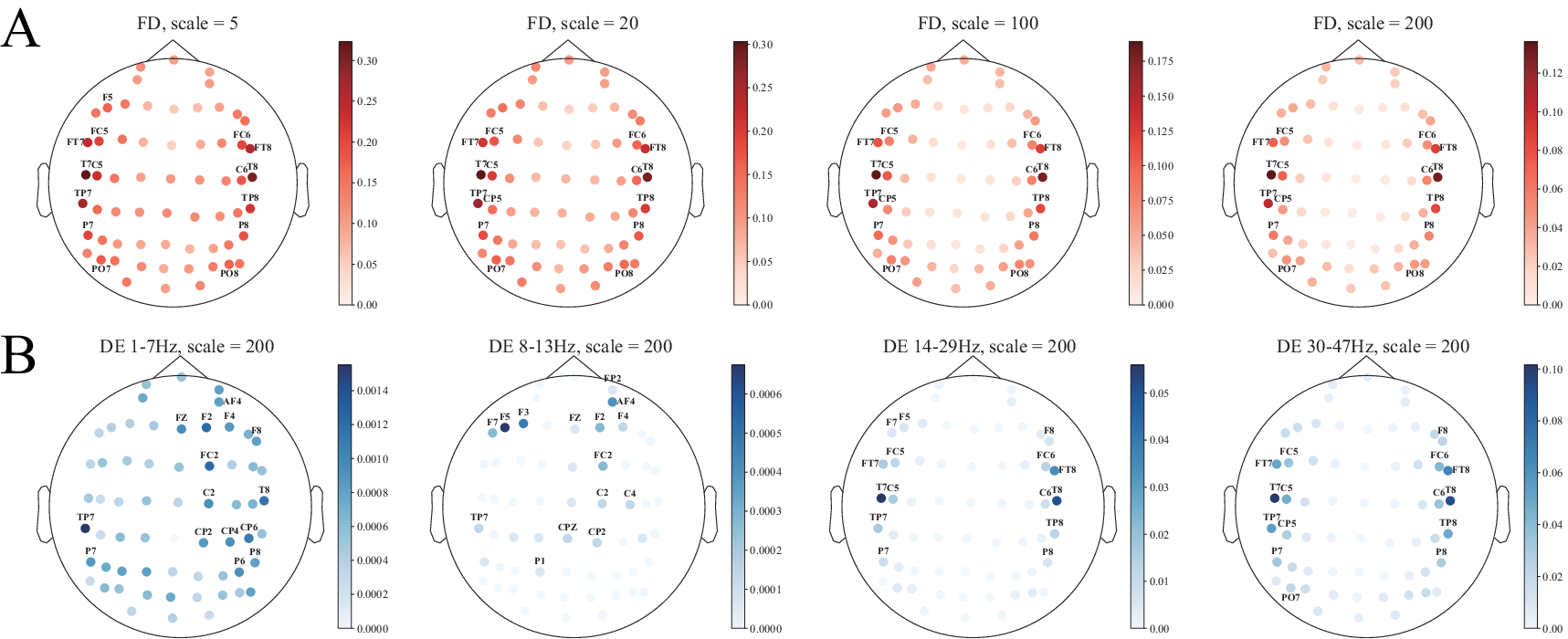}
  \caption{\textbf{Heterogeneous synchronized performance of electrodes.} \textbf{A} and \textbf{B} illustrate the synchronized percentage of each electrode with overall correlation on different scales of FD feature or different bands of DE feature. After Family-Wise Error Rate (FWER) Bonferroni correction for multiple comparison across films, pairwise subjects, and channels three dimensions, we calculate the percentage of significant correlations across films and pairwise subjects using the adjusted $p$ values of correlation ($p < 0.05$). 
  }
  \label{fig:overall_1}
\end{figure}

The study employs a user-dependent training strategy, configuring one classifier per subject. Each EEG record, approximately four minutes in duration, is segmented into 1-second samples from which features are extracted. A five-fold cross-validation approach is utilized for training and testing the classifiers, with data shuffling prior to model fitting to enhance performance. Hyperparameter optimization is conducted through nested grid search, with the SVM's gamma and C parameters ranging logarithmically from $2^{-4}$ to $2^{4}$, each fold independently selecting optimal parameters.

Results, summarized in Table \ref{tab:FeatureExtraction}, reveal that the FD feature achieved the highest classification accuracy at 87.96\%, surpassing other features significantly, with a notable 5\% improvement over the frequency domain feature, DE. It is important to note, however, that certain features may exhibit enhanced performance with alternative classifiers instead of the linear kernel SVM, such as those employing non-linear kernels. In brief, these results underscore the potential of FD and DE features in effectively capturing nuanced emotional information from EEG data.

\begin{figure}[htbp]
  \centering
  \includegraphics[width=0.7\linewidth]{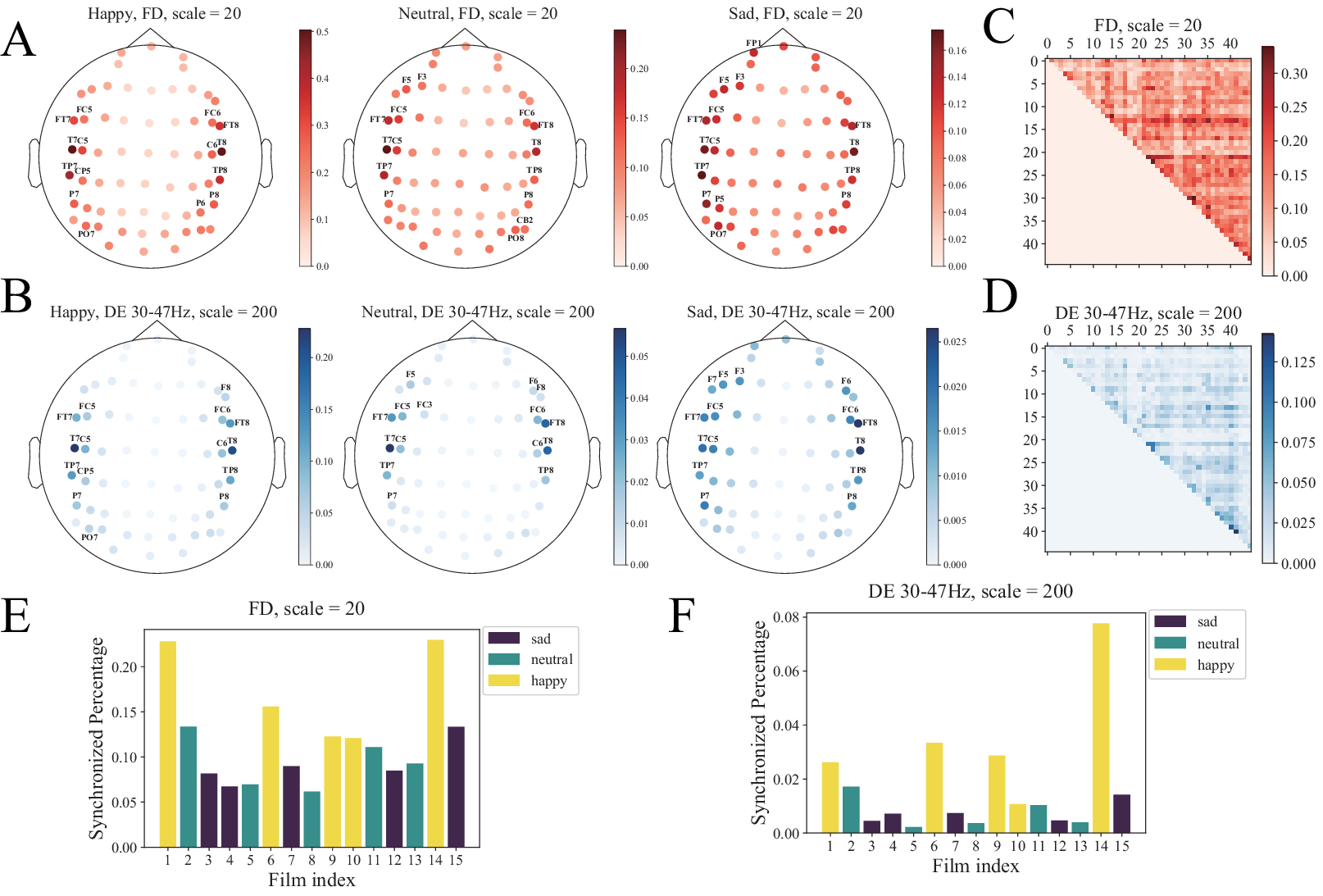}
  \caption{\textbf{Additional views of synchronized percentage for overall correlation}. \textbf{A} and \textbf{B} show the montage views of three emotion valences on FD, scale=20 and DE 30-47Hz, scale=200 respectively. \textbf{C} and \textbf{D} depict the synchronized percentage on each pairwise subject. \textbf{E} and \textbf{F} plot the synchronized percentage for each film.}
  \label{fig:overall_2}
\end{figure}

\subsection{Heterogeneous synchronized percentage of electrodes on overall ISC}

We apply the overall correlation to EEG features from identical electrodes, assessing their synchrony to preliminarily evaluate the viability of single-electrode EEG-based Inter-Subject Correlation (ISC). First-order Difference (FD) and Differential Entropy (DE) features are employed to compute ISC. Our findings reveal that electrodes near the temporal lobe exhibit the stronger ability to capture global synchrony, with consistent performance across various scales of the FD feature and different frequency bands of the DE feature.

For each film clip and each pair of subjects, we correlate two complete feature vectors from the same channel to obtain a 
$p$ value. This process results in a multi-dimensional $p$-value tensor ($F \times P \times C$), where $F$ denotes films, $P$ indicates pairwise subjects, and $C$ represents channels. Each scale of FD features and each band of DE features correspond to a specific $p$-value tensor. We apply a Family-Wise Error Rate (FWER) Bonferroni correction to the $p$-value tensor to address multiple comparisons. Following Hasson et al. \cite{hasson_intersubject_2004}, we use the 'synchronized percentage' as a metric for synchronization performance, calculated as the proportion of significant correlations ($p<0.05$) across channels, pairwise subjects, or films. This metric reflects the intensity of shared cognitive activity.

Our analysis across all scales of FD features and the $\beta, \gamma$ bands of DE features shows a heterogeneous yet consistent spatial pattern among the electrodes (Figure \ref{fig:overall_1}-A B). Electrodes near the bilateral temporal lobes are notably effective in capturing shared information processing related to emotions, a finding corroborated by other studies in EEG-based emotion recognition \cite{vytal_neuroimaging_2010}, \cite{zhang_visual--eeg_2022}, \cite{zheng_identifying_2019}. Similar patterns are observed for films with different emotional valences (Figure \ref{fig:overall_2}-A B), underscoring the consistency of these spatial patterns.  

Moreover, films with a 'happy' emotional valence tend to evoke stronger shared neural synchrony, suggesting a greater potential of such films to engage audiences similarly (Figure \ref{fig:overall_2}-A B). Additionally, both FD and DE features show synchronization across many pairwise subjects, confirming that the observed overall synchronization is not limited to a few pairs but is rather widespread (Figure \ref{fig:overall_2}-C D). The consistent induction of neural synchrony by each film, albeit with varying potentials linked to different valences, further supports our findings (Figure \ref{fig:overall_2}-E F).
Besides, the montage view plots are improved based on \href{https://mne.tools/0.21/generated/mne.channels.DigMontage.html?highlight=montage%20plo#mne.channels.DigMontage.plot}{mne.channels.DigMontage.plot} \cite{gramfort_meg_2013}.

\begin{figure}[htbp]
  \centering
  \includegraphics[width=0.7\linewidth]{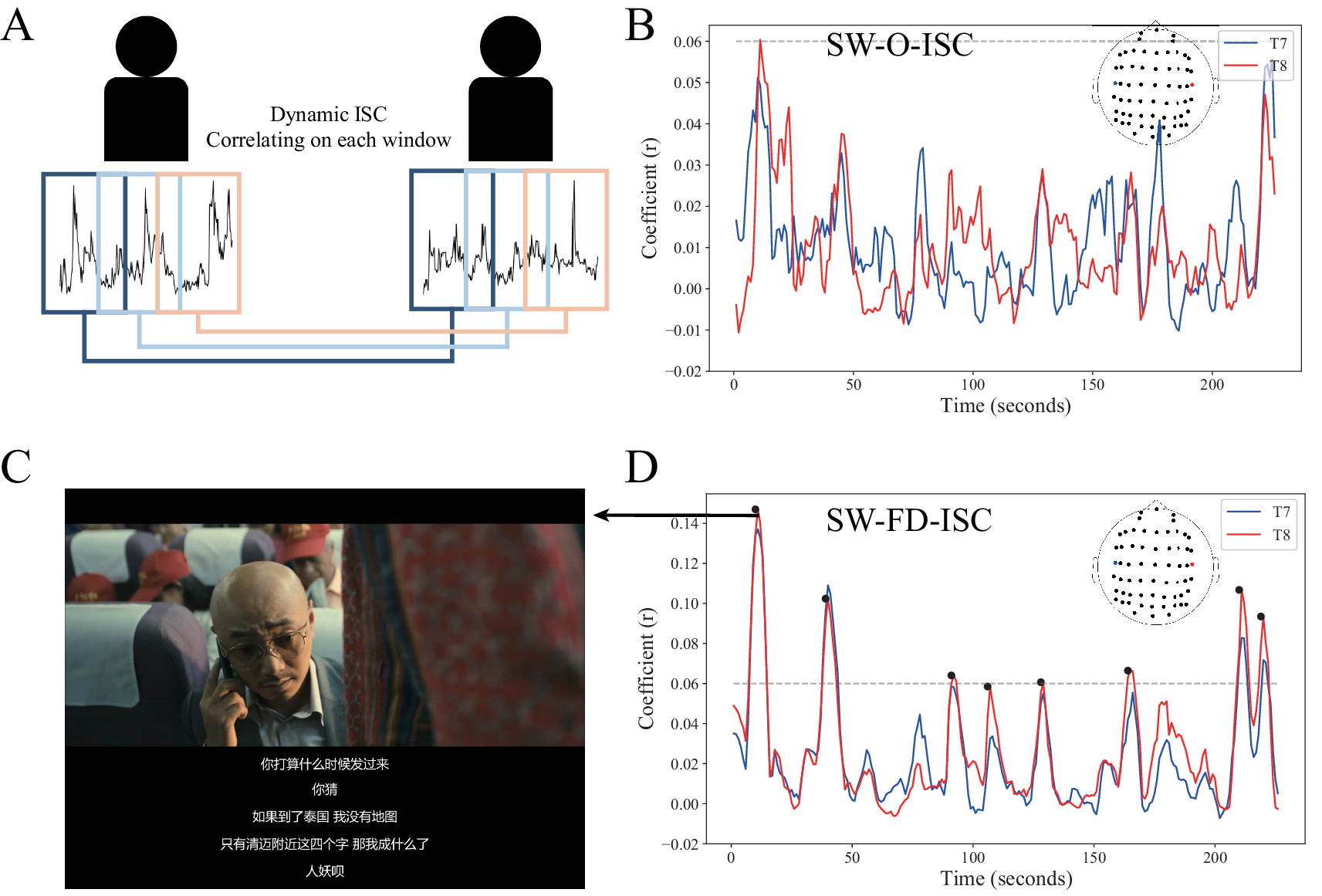}
  \caption{\textbf{Case comparison of the dynamic ISCs at electrodes T7 and T8.} \textbf{A.} The illustration of computing one dynamic ISC between one pairwise subject. \textbf{B} and \textbf{D} The example dynamic ISCs of SW-O-ISC and SW-FD-ISC on the electrodes T7,T8 for the first film clip in SEED. \textbf{C.} A scene from the first film clip corresponding to the window at the first peak.}
  \label{fig:dynamic_1}
\end{figure}

\subsection{Validating the reliability of dynamic ISC}

\begin{figure}[htbp]
  \centering
  \includegraphics[width=0.9\linewidth]{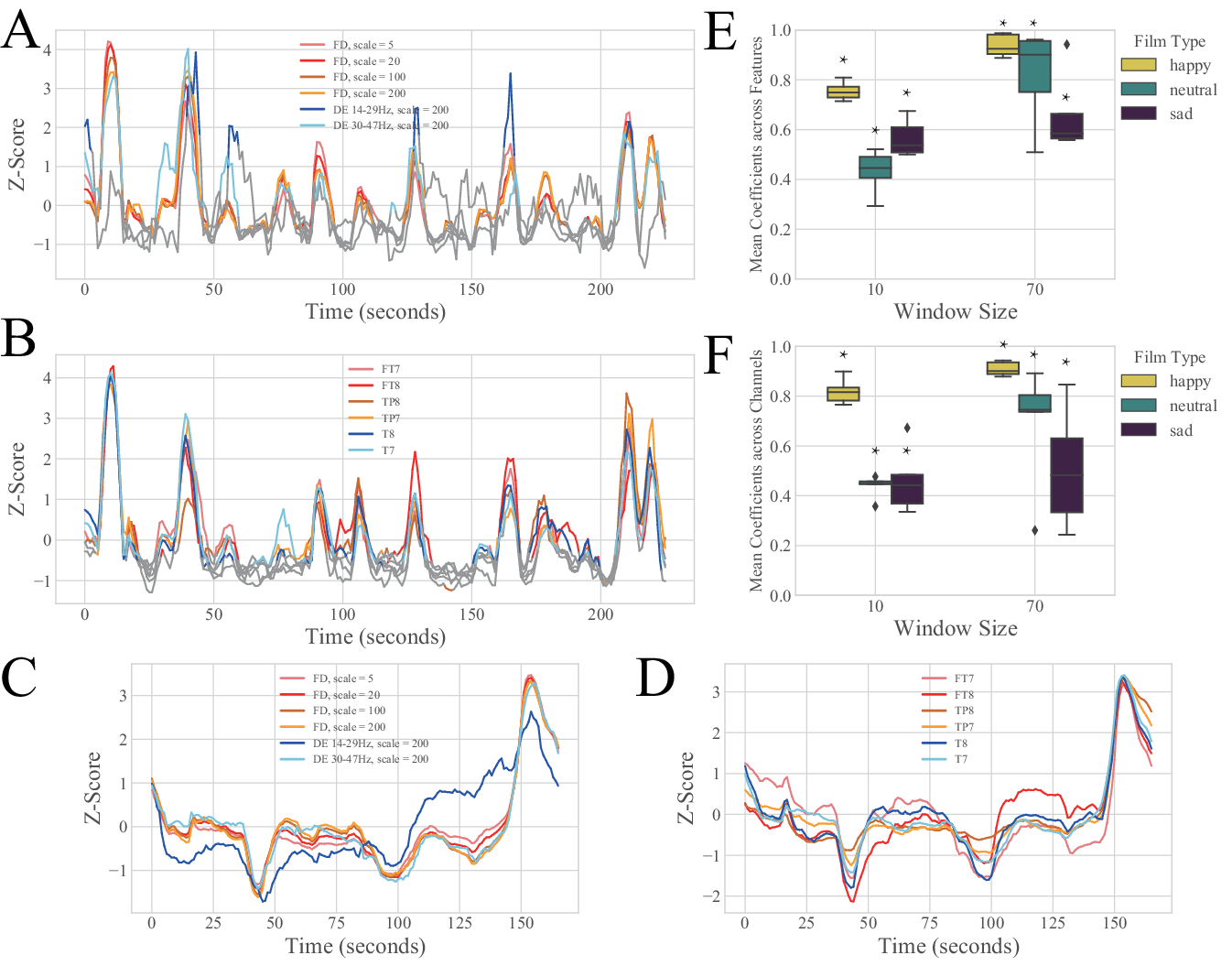}
  \caption{\textbf{The consistent dynamic ISCs on both 10 and 70 window sizes.} \textbf{A} and \textbf{C} show the dynamic ISCs across features on the electrode T7, window size 10s or 70s respectively. \textbf{B} and \textbf{D} illustrate the dynamic ISCs across key electrodes, window size 10s or 70s respectively. The significance of coefficients in each dynamic ISC is assessed using the Wilcoxon signed-rank test with adjustments for multiple comparisons via the Benjamini-Hochberg False Discovery Rate (FDR). Only significant coefficients are color-coded. \textbf{E} and \textbf{F} depict the mean correlation coefficients of dynamic ISCs across features (with the electrode T7) or channels (with FD, scale=20) under category view. Results with other configurations have similar performance and are not shown for brevity. The significance of each category is tested using one-sample t-tests against a threshold of 0.2. }
  \label{fig:dynamic_2}
\end{figure}

In this section, we employ the sliding window correlation technique (Figure \ref{fig:dynamic_1}-A) to analyze the dynamic shared neural responses across participants at critical electrodes (FT7, FT8, TP7, TP8, T7, T8). We calculate the mean correlation coefficients derived from the dynamic inter-subject correlation (ISC) of emotion-related features or key electrodes to assess the reliability of our methods. Our findings reveal that the similarity of different dynamic ISCs exhibits significant performance across all film categories and both window sizes, particularly for 'happy' films. This suggests that consistent dynamic synchronous neural responses to stimuli can be effectively captured through single-electrode sliding window correlations, underscoring the potential of this method to represent group-level emotional arousal triggered by the films.

We begin with a comparative analysis of the dynamic ISC at electrodes T7 and T8 using sliding window correlations with original inter-subject correlation (SW-O-ISC) and first-difference inter-subject correlation (SW-FD-ISC) at a scale of 20, focusing on the first film from the SEED dataset (Figure \ref{fig:dynamic_1}-B D). The grey dotted lines in the plots indicate the highest correlation coefficient for SW-O-ISC. Notably, many instances in SW-FD-ISC exceed this benchmark. This comparison demonstrates that SW-FD-ISC provides smoother, stronger, and more consistent responses than SW-O-ISC, highlighting its efficacy in capturing emotion-related dynamic ISC at pivotal electrodes. Additionally, we illustrate a 10-second scene corresponding to the first peak in SW-FD-ISC, which coincides with a humorous moment eliciting significant emotional arousal in the audience (Figure \ref{fig:dynamic_1}-C). 

Further comprehensive comparisons of dynamic ISCs across different emotion-related features or several key electrodes are conducted. In each film clip, various scales of first-difference (FD) features and the $\beta$ or $\gamma$ bands of differential entropy (DE) features are employed to extract population-level dynamic synchronous responses at each key electrode. The significance of coefficients in each dynamic ISC is assessed using the Wilcoxon signed-rank test with adjustments for multiple comparisons via the Benjamini-Hochberg False Discovery Rate (FDR) \cite{benjamini_discovering_2010}. Only significant coefficients are color-coded, and Z-score normalization is employed prior to comparing the performance of dynamic ISCs. In the first film, significant peaks are consistently observed in dynamic ISCs at window 10, indicating the reliable capture of fine-grained and emotion-related information (Figure \ref{fig:dynamic_2}-A B). Moreover, dynamic ISCs at window 70 consistently capture all significant variations in shared responses, suggesting the presence of similar cognitive activities among the audience on a larger scale (Figure \ref{fig:dynamic_2}-C D).

To assess the performance across all film clips, we calculate the mean correlation coefficients for each individual film clip, categorizing them by movie emotion labels, each category containing five clips. These coefficients are derived by averaging the pairwise correlations of dynamic ISCs across various features or key channels. We evaluate the significance of these means using one-sample t-tests against a threshold of 0.2. The resulting box plots for each category demonstrate significant performance, affirming that the similarity among dynamic ISCs for film clips within each category is substantial (Figure \ref{fig:dynamic_2}-E F). Particularly noteworthy are films in the 'happy' category, which display mean coefficients approaching 0.8. This suggests that these films are exceptionally effective at engaging the audience and eliciting synchronized neural responses, indicative of strong emotional engagement.

\section{Discussion and Conclusion}
In this study, we explore the viability of using EEG-based inter-subject correlation (ISC) for automatic and unconscious emotion annotation. With a particular focus on reducing the number of electrodes required, we have developed a novel single-electrode and feature-based dynamic ISC method. The contributions of our work are threefold: (1) We reidentify two effective emotion features, one from the time domain and one from the frequency domain: first-order difference (FD) and differential entropy (DE). (2) We utilize the overall correlation to demonstrate the heterogeneous synchronized performance of electrodes. This performance is consistent with neural emotion patterns identified in previous research, providing a validation of our method's effectiveness. (3) Applying a sliding window correlation, we illustrate the similarity of dynamic ISCs across various features or key electrodes for each film clip. This indicates the reliability of our method in capturing consistent dynamic shared neural synchrony among individuals induced by evocative film clips, highlighting its potential as an indicator of human continuous emotion arousal.

Our findings on the temporal lobe's dominance in neural synchrony are consistent with a breadth of multidisciplinary literature, encompassing neuroscience and affective computing. Neuroscientists, such as Vytal and Hamann \cite{vytal_neuroimaging_2010}, using functional Magnetic Resonance Imaging (fMRI), identified distinct activation patterns associated with basic emotions in specific brain regions: happiness primarily activates the right superior temporal gyrus (STG) and left anterior cingulate cortex (ACC), while sadness is primarily associated with the left medial frontal gyrus (medFG). Furthermore, several studies have employed EEG and computational models to delineate spatial distribution patterns of emotional responses. Critical emotion patterns located at the temporal lobe have been observed through various methods, including DE feature frequency bands, weight distributions of deep belief networks \cite{wei-long_zheng_investigating_2015}, average energy distributions \cite{zhao_classification_2019}, \cite{zheng_identifying_2019}, spatio-temporal feature characteristics by contrastive learning intracranial seizure analysis (CLISA) \cite{shen_contrastive_2022}, and gradient visualization of a teacher-student model \cite{zhang_visual--eeg_2022}. These findings collectively underscore the integral role of the temporal lobes in brain emotion functions. The heterogeneous electrode performance observed in our results aligns with these established spatial patterns of emotions.

\begin{credits}
\subsubsection{\ackname} This work is supported by National Key Research and Development Program of China (2021YFB2700300), Program of National Natural Science Foundation of China (62141605, 12201026,12301305)

\subsubsection{\discintname}
The authors have no competing interests. 
\end{credits}
%
%
%
%
\bibliographystyle{splncs04}
\bibliography{Zotero}
\end{document}